\documentclass[sn-mathphys-num]{sn-jnl} 

\usepackage{graphicx}%
\usepackage{multirow}%
\usepackage{amsmath,amssymb,amsfonts}%
\usepackage{amsthm}%
\usepackage{mathrsfs}%
\usepackage[title]{appendix}%
\usepackage{xcolor}%
\usepackage{textcomp}%
\usepackage{manyfoot}%
\usepackage{booktabs}%
\usepackage{algorithm}%
\usepackage{algorithmicx}%
\usepackage{algpseudocode}%
\usepackage{listings}%

\begin{document}

\title[Predictability of temporal network dynamics in normal ageing and brain pathology]{Predictability of temporal network dynamics in normal ageing and brain pathology}

\author[1]{\fnm{Annalisa} \sur{Caligiuri}}
\author[2, 3]{\fnm{David} \sur{Papo}}
\author[4, 5]{\fnm{G\"orsev} \sur{Yener}}
\author[6, 7]{\fnm{Bahar} \sur{G\"untekin}}
\author[1]{\fnm{Tobias} \sur{Galla}}
\author[1]{\fnm{Lucas} \sur{Lacasa}}
\author*[1]{\fnm{Massimiliano} \sur{Zanin}}\email{mzanin@ifisc.uib-csic.es}

\affil[1]{ \orgname{Institute for Cross-Disciplinary Physics and Complex Systems IFISC (CSIC-UIB)}, \state{Palma de Mallorca}, \country{Spain} }

\affil[2]{ \orgdiv{Department of Neuroscience and Rehabilitation}, \orgname{University of Ferrara}, \state{Ferrara}, \country{Italy} }

\affil[3]{ \orgdiv{Center for Translational Neurophysiology}, \orgname{Fondazione Istituto Italiano di Tecnologia}, \state{Ferrara}, \country{Italy} }

\affil[4]{ \orgdiv{Department of Neurology, Faculty of Medicine}, \orgname{Dokuz Eyl\"ul University}, \state{Izmir}, \country{T\"urkiye} }

\affil[5]{ \orgname{IBG: International Biomedicine and Genome Center}, \state{Izmir}, \country{T\"urkiye} }

\affil[6]{ \orgdiv{Neuroscience Research Center, Research Institute for Health Sciences and Technologies (SABITA)}, \orgname{Istanbul Medipol University}, \state{Istanbul}, \country{T\"urkiye} }

\affil[7]{ \orgdiv{Department of Biophysics, School of Medicine}, \orgname{Istanbul Medipol University}, \state{Istanbul}, \country{T\"urkiye} }

\abstract{Spontaneous brain activity generically displays transient spatiotemporal coherent structures, which can selectively be affected in various neurological and psychiatric pathologies. Here we model the full brain's electroencephalographic activity as a high-dimensional functional network performing a trajectory in a latent graph phase space. This approach allows us to investigate the orbital stability of brain's activity and in particular its short-term predictability. We do this by constructing a non-parametric statistic quantifying the expansion of initially close functional network trajectories. We apply the method to cohorts of healthy ageing individuals, and patients previously diagnosed with Parkinson's or Alzheimer's disease. Results not only characterise brain dynamics from a new angle, but further show that functional network predictability varies in a marked scale-dependent way across healthy controls and patient groups. The path towards both pathologies is markedly different. 
Furthermore, healthy ageing's predictability appears to strongly differ from that of Parkinson's disease, but much less from that of patients with Alzheimer's disease.}

\keywords{Functional networks, Predictability, Parkinson's disease, Alzheimer's disease}

\maketitle

\section{Introduction}

Modelling and quantifying brain activity, and how this activity is modified in neurological and psychiatric pathology, are daunting tasks. Probing brain dynamics by perturbing it in various ways (pharmacological, via electrical brain stimulation, or by cognitive testing) has been the standard method in neuroscience for a very long time. At the same time, it is now clear that ongoing (unperturbed) brain activity is far from random \cite{arieli1996dynamics}, and contains information which can be used to characterise both healthy and pathological brain dynamics \cite{papo2014functional}.

Brain activity has traditionally been modelled as a spatially-extended nonlinear dynamical system with both deterministic and random characteristics \cite{gao2012multiscale}. At the same time, brain disorders can be thought of as dynamical regimes, emerging as system control parameter variations induce qualitative changes in collective variables describing neural activity \cite{mackey1977oscillation, glass1979pathological, pezard1996depression, pezard1998entropy, glass2015dynamical}. Following this view, brain disorders have been studied using methods from nonlinear dynamics and time series analysis \cite{abarbanel1993analysis, hegger1999practical, kantz2003nonlinear, rezek1998stochastic, zou2019complex, sannino2017visibility}. These ideas have contributed to characterising both healthy and pathological brain activity \cite{molnar1999brain, pezard2001investigation, lehnertz2001nonlinear, stam2005nonlinear, sannino2017visibility}. They allow complementary and often non-invasive quantitative diagnosis, and provide clues for further understanding the physiological mechanisms behind altered brain behaviour. Here we aim to quantify the predictability of spatially-extended dynamical systems, acknowledging the multiplicity of dynamical scales. We will now explain in turn what we mean by this.

\medskip
\noindent {\it Predictability --} The lack of predictability refers to the extent to which uncertainty on a system's initial condition propagates and is amplified over time \cite{abarbanel1990prediction, boffetta2002predictability}. Characterising predictability is an established problem in the theory of dynamical systems \cite{eckmann1985ergodic, boffetta2002predictability, kantz2003nonlinear}. 
Standard tools from the theory of nonlinear and chaotic systems are available to quantify error growth and information production rates \cite{Shaw+1981+80+112}, reflecting the speed at which predictability is lost.  In predictable systems, small initial deviations are usually either bounded or shrink over time, while chaotic systems are affected by sensitivity to initial conditions, so that small initial deviations are often amplified exponentially. 
A popular quantifier is the rate of separation of initially close trajectories. The so-called largest Lyapunov exponent (LLE) captures such exponential rate, such that in a system with a positive LLE, even tiny errors in the initial conditions grow exponentially, making long-term prediction impossible \cite{kantz2003nonlinear}.
Alternative measures have been also introduced to characterise the non-exponential growth of uncertainty; for example, in the case of weak chaos, the generalised Lyapunov exponent assumes a non-zero value even if the growth is weaker than exponential \cite{klages2013weak,zaslavsky1992weak}.

\medskip \noindent 
{\it Multiple timescales in brain dynamics --} An accurate characterisation of spontaneous brain activity also requires adequately considering its multiscale nature. Indeed, as most complex many-body biological systems operating far from equilibrium, spontaneous brain activity is known to possess temporal structure in a wide range of time scales, and possibly some relationship among the dynamics on these different time scales. At the same time, statistical, dynamical and thermodynamical properties seem to be scale-dependent \cite{egolf2000equilibrium, battle2016broken, tan2021scale, bernardi2023time}. For instance, while at long timescales the presence of non-trivial fluctuations typical of complex systems is associated with temporally ordered structure \cite{novikov1997scale, linkenkaer2001long, beggs2003neuronal, bianco2007brain, freyer2009bistability, allegrini2010fractal}, at short time scales activity is associated with exponential relaxation, where memory of the system's past state is quickly lost. Another documented scale-dependent property of the brain is that of time irreversibility, which may differ across scales or frequencies \cite{zanin2020time}. In a nutshell, any brain-dynamical property should be assessed in a scale-dependent fashion \cite{gao2012multiscale}.

\medskip 
Finally, in addition to the above-mentioned dynamical properties, brain activity also shows complex {\it spatial} structure. As a matter of fact, over the past few years it has become standard to understand the form of strong disorder of the brain's spatial structure in terms of a network structure \cite{bullmore2009complex, papo2023does}.
Altogether, the brain can effectively be thought of as a {\it networked} dynamical system. Its dynamical regimes and transitions between different types of dynamics are captured by changes in the network topology \cite{thompson2017static}.
The brain's activity can therefore be modelled as a graph evolving over time, i.e. as a {\it network trajectory} \cite{lacasa2022correlations}. This interpretation aligns with a recent reinterpretation of temporal networks \cite{holme2012temporal, masuda2016guide}, whereby time series of networks represent a trajectory of a latent graph dynamical system \cite{lacasa2022correlations, caligiuri2023lyapunov, danovski2024dynamical, caligiuri2024characterising}. Insight into the brain's dynamics can therefore be gained using the tools of time series analysis and signal processing \cite{lacasa2022correlations, caligiuri2023lyapunov, bauza2023characterization, andres2024detecting, williams2022shape, gallo2024higher, lacasa2024scalar}. Quantifying brain's predictability thus unavoidably requires the use specific, network-based estimators defined in the context of temporal network trajectories \cite{caligiuri2023lyapunov, danovski2024dynamical, caligiuri2024characterising} over multiple timescales of activity. We note that dynamical metrics such as Lyapunov exponents have been previously measured in brain activity data such as that from electroencepalogram (EEG) or magnetoencephalogram (MEG) measurement. This type of analysis has been carried out for patients with various psychiatric conditions, including depression \cite{roschke1994nonlinear, kustubayeva2021lyapunov}, schizophrenia \cite{roschke1994nonlinear, kim2000estimation, singh2023cognitive}, Parkinson's disease (PD) \cite{saikia2019significance}, and Alzheimer's disease (AD) \cite{jeong2002nonlinear, stam2005nonlinear, dauwels2010diagnosis}. However, most of these studies (i) did not explicitly take into account an underlying spatial network structure, and (ii) concentrated on single time scales (or, equivalently, on a global frequency range \cite{takahashi2013complexity}).

\medskip 
In this study we aim to assess if and how the predictability of spontaneous brain activity is modulated in healthy human ageing and in two different types of dementia, AD and PD, as well as in their respective precursor conditions. To address this issue, we initially reconstruct network trajectories from brain activity data of the abovementioned cohorts of controls and patients. We then 
introduce a measure for temporal network predictability, conceptually close to the network Lyapunov exponent proposed in \cite{caligiuri2023lyapunov, danovski2024dynamical}, but applicable in the more general case where there is no prior assumption of any particular functional shape --exponential or otherwise-- for the expansion rate of nearby trajectories. This measure is designed to characterise the predictability of brain network dynamics at different timescales. Building on this measure, we are able to identify simple, interpretable features which indicate that, as compared to the control group, predictability is altered both in AD and PD in a statistically significant manner. We also observe that these differences show non-trivial frequency-dependence. We relate these results to concepts previously presented in the literature, including the irreversibility of brain dynamics and the stability of specific network patterns.

\section{Materials and methods}

\subsection{Control subjects and patients recruiting and selection}

This study included ninety-eight subjects, as also analysed in previous works \cite{zanin2022fast, zanin2022telling}. The subjects are divided into six different groups: healthy (control) young, healthy (control) elderly, patients diagnosed with Mild Cognitive Impairment (MCI), patients diagnosed with Alzheimer's disease with dementia (AD), patients diagnosed with Parkinson's disease with mild cognitive impairment (PD-MCI), and patients diagnosed with Parkinson's disease with dementia (PD-D).
Demographic information of the participants is presented in Tab. \ref{tab:subjData}.

Neurologists at Istanbul Medipol University and Dokuz Eyl\"ul University performed a clinical diagnosis of the patients, and all patients underwent complete neurological structural magnetic resonance imaging (MRI). An extensive battery of neuropsychological tests was performed for patient groups and elderly subjects. Patients with amnesic MCI were diagnosed according to the NIA-AA criteria \cite{albert2011diagnosis}; those with PD-MCI, according to the Movement Disorder Society (MDS) Level 2 criteria \cite{litvan2012diagnostic}. Probable PD-D diagnosis was made according to the Movement Disorder Society (MDS) Level 1 criteria \cite{emre2007clinical, dubois2007diagnostic}. All patients with AD were diagnosed according to the National Institute of Aging-Alzheimer's Association diagnostic guidelines \cite{mckhann2011diagnosis}. All participants with PD-MCI and PD-D were using a dopaminergic medication, including levodopa and/or dopamine agonist and/or monoamineoxidase B (MAO-B) inhibitor. All AD patients were on medication with cholinesterase inhibitors, and half of them were also on memantine, a NMDA receptor antagonist. 
The healthy elderly volunteers included in this study had no neurological abnormality or global cognitive impairment (Mini-mental State Examination (MMSE) score $\geq 27$). Inclusion criteria for amnesic MCI included: i) leading an independent life in the community, ii) memory problem as defined by performance $\geq 1.5$ standard deviations below that of age and education-matched controls in a set of neuropsychological tests, and iii) no impairment of daily living activities, clinical dementia rating (CDR) score of 0.5. Young healthy controls' inclusion criteria were no presence or history of neurological and psychiatric abnormalities, of drug and alcohol abuse and/or the Mini-mental State Examination (MMSE) score of $\geq 27$.

Exclusion criteria for all participants were: i) history of neurological and/or psychiatric abnormalities, including evidence of depression as demonstrated by Yesavage Geriatric Depression Scale scores higher than 13 \cite{demet2002depressive}; ii) presence of non-stabilised medical illnesses; iii) history of severe head injury and alcohol or drug misuse; iv) use of psychoactive drugs or cognitive enhancers, including acetylcholinesterase inhibitors; v) presence of vascular brain lesions, brain tumour as per MRI, and hydrocephalus.

The study conformed to the Declaration of Helsinki's principles. All participants and/or their relatives provided informed consent for the study, which was approved by the local ethical committee (Istanbul Medipol University Ethical Committee, Report No: 10840098-604.01.01-E.8374 and Dokuz Eyl\"ul University Ethical Committee, 01.03.2018/3821 GOA and 2018.KB.SAG.084).

\begin{table}[h]
\centering
\begin{tabular}{|l|l|p{2.4cm}|p{2.2cm}|p{3.3cm}|}
\hline
{\bf Subject group} & {\bf Size} & {\bf Of which \newline men/women} & {\bf Avg. age \newline (std.)} & {\bf Avg. years of \newline education (std.)} \\ \hline
Control (young) & $18$ & $9$/$9$ & $24.1$ ($3.68$) & $15.5$ ($1.54$) \\ \hline
Control (elderly) & $19$ & $11$/$8$ & $69.1$ ($7.25$) & $10.9$ ($4.67$) \\ \hline
MCI & $16$ & $7$/$9$ & $70.4$ ($5.05$) & $9.4$ ($5.88$) \\ \hline
AD & $19$ & $5$/$14$ & $73.2$ ($5.68$) & $8.8$ ($4.40$) \\ \hline
PD-MCI & $14$ & $9$/$5$ & $71.1$ ($6.63$) & $11.4$ ($5.18$) \\ \hline
PD-D & $12$ & $9$/$3$ & $73.0$ ($6.71$) & $5.9$ ($5.21$) \\ \hline
\end{tabular}
\caption{Demographic data of the participants comprising the data set used in this study.\label{tab:subjData}}
\end{table}

\subsection{Electroencephalographic data recording}

EEG activity was recorded with a Brain Products BrainAmp 32 Channel DC system (band limits: $0.001$-$250$ Hz; sampling rate: $500$ Hz). The EasyCap with 32 Ag/AgCl electrode montage device with electrodes placed according to the 10/20 system was used. All electrodes' impedances were kept below $10$ k$\Omega$. A1 and A2 electrodes placed on the earlobes were used as reference. The electrooculogram was recorded via two Ag/AgCl electrodes placed on the left eye's medial upper and lateral orbital rim. The EEG was recorded in a dimly isolated room in two different centres with identical recording equipment, and the EEG recording procedure was applied precisely in the two centres, Istanbul Medipol University's SABITA Neuroscience Research Center EEG laboratory, and Dokuz Eyl\"ul University's Multidisciplinary Brain Dynamics Research Center. For all participants, EEG activity was recorded during 4 minutes in the eyes-open condition and 4 minutes in eyes-closed condition, yielding approximately $240,000$ data points per channel and subject. In the former condition, participants were asked to look at the black screen. Participants were monitored with a video camera throughout the EEG recording session.

The resulting time series were manually inspected for evident errors; an automatic artefact removal procedure was further applied, based on subtracting the two main components of all time series using Independent Component Analysis.

\subsection{Functional network trajectory reconstruction and distance function}

\begin{center}
\begin{figure}
    \rotatebox{90}{\includegraphics[width=0.63\linewidth]{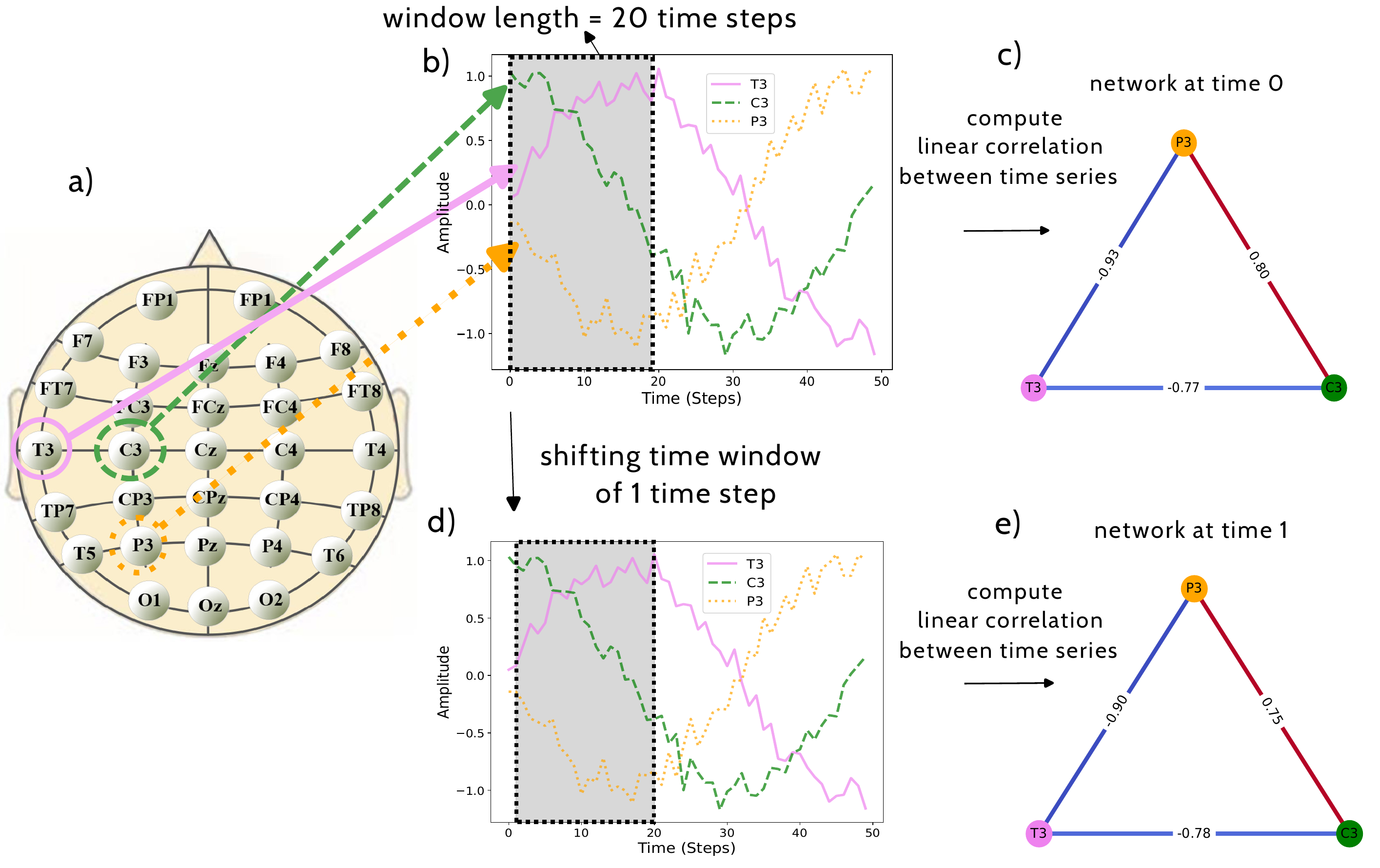}}
    \caption{Example of the functional brain network reconstruction procedure. We depict the procedure for the simplified case of three nodes (i.e. three electrodes) and the same number of time series (see panel \textit{a}). In the second step (panel \textit{b}), we consider the time series of the amplitude of the signal and select a time window (grey shaded region) with a length of 20 time steps. Within it, we compute the linear correlation between pairs of the three time series, resulting in correlation values that represent the edge weights in the network at time \(t=0\) (see panel \textit{c}). 
    To obtain the network at the following time step, we shift the time window by 1 time step (2 milliseconds), as shown in panel \textit{d}), and repeat the procedure to construct the network at time \(t=1\) (panel \textit{e}). Iterating this procedure, we obtain the full functional network trajectory.}
    \label{fig:Cartoon}
\end{figure}
\end{center}

Functional networks representing brain dynamics were reconstructed following standard approaches from the literature \cite{bullmore2009complex, papo2023does}. Brain activity was recorded using $N=30$ brain electrodes, resulting in networks with $N$ nodes, where each node represents a distinct time series corresponding to the signal detected at the electrode location.
To enable more robust statistical analyses, each recording was divided into 20 consecutive, non-overlapping time segments. Each segment includes simultaneous signals from all 30 electrodes and is treated as an independent virtual recording. These segments are referred to as ``trials.'' Thus, a trial consists of the full set of 30 time series extracted from one segment of the recording and provides the basis for constructing a corresponding network trajectory.
The process for generating a network trajectory from each trial is as follows:
\begin{enumerate}
    \item \textit{Correlation calculation:} The signals from the 30 electrodes are analysed over a defined time interval, with a duration determined by the \textit{window length}. For each pair of nodes (electrodes), we compute the Pearson's linear correlation coefficient between the time series within that interval.
    \item \textit{Network reconstruction:} The resulting pairwise correlations form a weighted, symmetric adjacency matrix, where each entry represents the absolute value of the linear correlation between the corresponding electrode signals. The weights capture the strength of the functional connectivity between nodes.
    \item \textit{Dynamic evolution:} The analysis window is then shifted by 1 time step, corresponding to 2 milliseconds, and the correlation computation is repeated to generate the next network snapshot. Repeating this process throughout the whole trail yields the full trajectory, composed of $5,000$ functional networks, capturing the temporal evolution of the brain's functional connectivity.
\end{enumerate}
This procedure is illustrated in the cartoon in Fig. \ref{fig:Cartoon}.
In total, we analysed 1,960 trials, categorised as follows: 360 from young control participants, 380 from elderly controls, 320 from MCI (potential AD) patients, 380 from AD patients, 280 from PD-MCI patients, and 240 from PD-D patients.

\medskip
In order to compare the evolution of two network trajectories, among other distance functions \cite{wills2020metrics}, here we define the distance between two arbitrary networks $\mathcal{A}, \mathcal{B}$ as:
\begin{equation}\label{eq:distance}
    d( \mathcal{A}, \mathcal{B} ) = \frac{2}{N (N - 1)} \sum _{i> j} | a_{i, j} - b_{i,j} |,
\end{equation}
where $\mathcal{A} = \{a_{ij}\}_{i,j=1}^N$ and $\mathcal{B} = \{b_{ij}\}_{i,j=1}^N$ are the two weighted adjacency matrices to be compared, $N$ is the number of nodes in each network (here $N = 30$), and $i$ and $j$ take values from $1$ to $N$. Each $a_{i,j}$ and $b_{i,j}$ can take values between zero and one. Consequently, $d$ takes values from $0$ (for pairs of identical networks) to $1$ (when $|a_{i,j} - b_{i,j}|=1 \ \forall i,j$, i.e, when each pair $i>j$ has full correlation in one network ($a_{i,j}=1$) and no correlation in the other ($b_{i,j}=0$). The value $d=1$ can only be reached in very artificial situations and is never observed in the experiments examined here. For illustration, sampling networks at random from the trajectories of different subjects and conditions yields an average distance of $\langle d \rangle  \approx 0.47$ between networks (standard deviation $0.07$).

\subsection{Quantifying network predictability: a non-parametric expansion rate of close trajectories}

\begin{figure}[tb!]
    \includegraphics[width=0.9\linewidth]{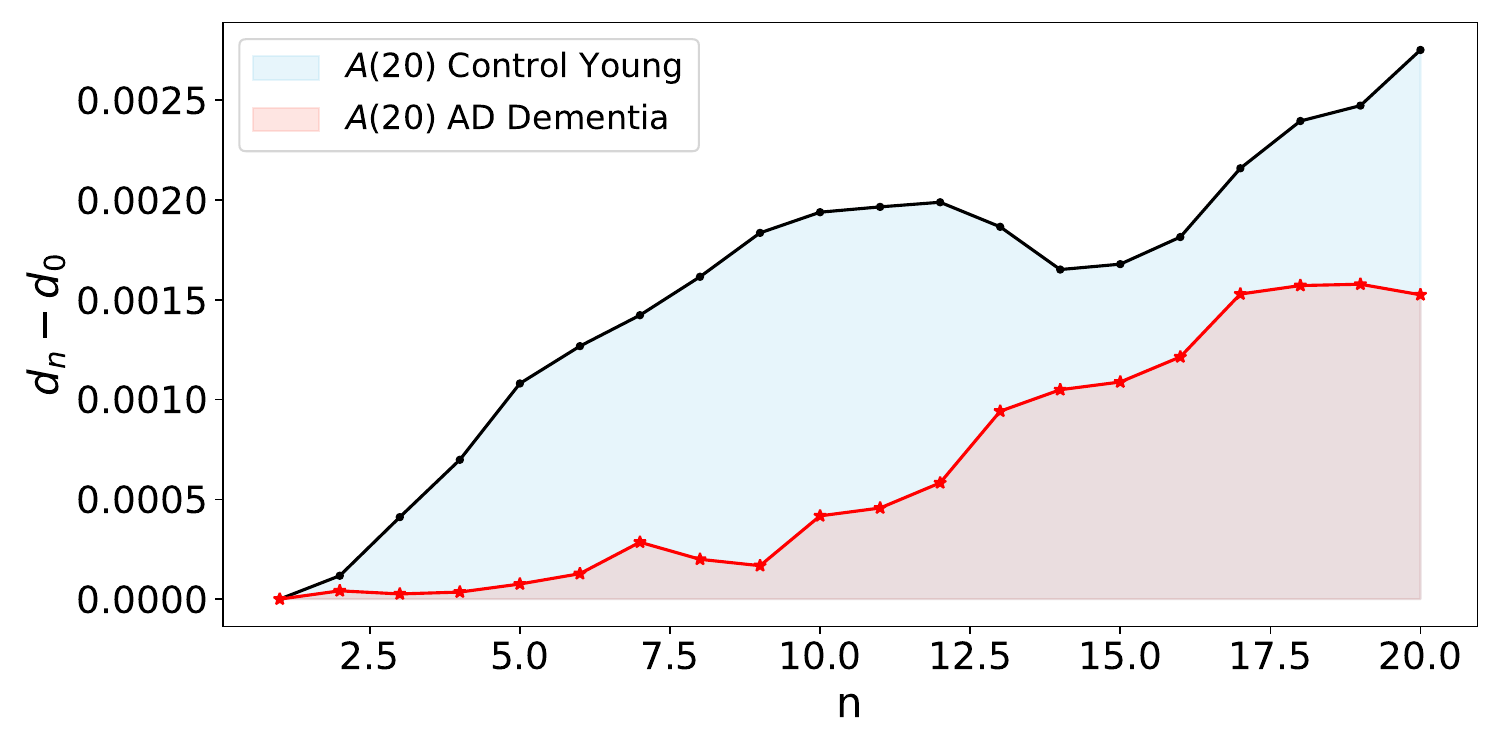}
    \caption{Illustration of the non-parametric expansion measure $\texttt{A}(n)$ for two different pairs of initially close network trajectories. The plot shows how $\texttt{A}(n)$ (Eq.\ref{eq:non_parametric}) captures the rate of expansion over time by representing the area below the distance curve between two initially close trajectories, after subtracting the initial distance $d_0$. In this figure, the area under the curve associated with the young control is larger than that of the AD Dementia patient, indicating that the former exhibits a faster expansion compared to the latter.}
    \label{fig:example_A}
\end{figure}

We now introduce a non-parametric way to measure the rate of expansion over time of initially close network trajectories. 
Suppose that two network trajectories are at distance $d_0$ at time $t=0$, and denote $d_t$ the distance after $t$ time steps. We define the non-parametric expansion $\texttt{A}(n)$ as:
\begin{equation}
   \texttt{A}(n) = \frac{1}{n}\sum^n_{t=1} (d_t-d_0),
   \label{eq:non_parametric}
\end{equation}
where $n$ is the total number of steps over which we are monitoring the expansion. Up to possible normalisation factors (e.g., length of the time step), the quantity $\texttt{A}(n)$ represents the area below the plot of $d_t$ as a function of $t$, after subtracting the initial value $d_0$. The larger  $\texttt{A}(n)$, the larger that area, and the faster the two initial close trajectories expand over time. An illustration of $\texttt{A}(n)$ for two different pairs of initially close trajectories, one expanding faster than the other, is shown in Fig. \ref{fig:example_A}. 

\begin{figure}
    \includegraphics[width=0.32\linewidth]{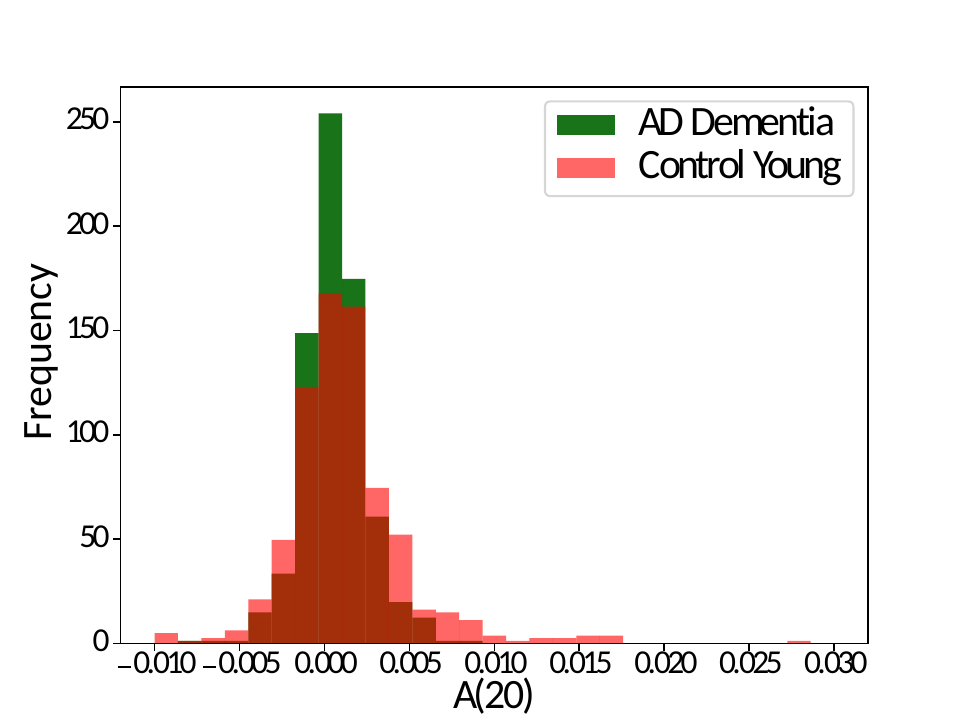}
        \includegraphics[width=0.32\linewidth]{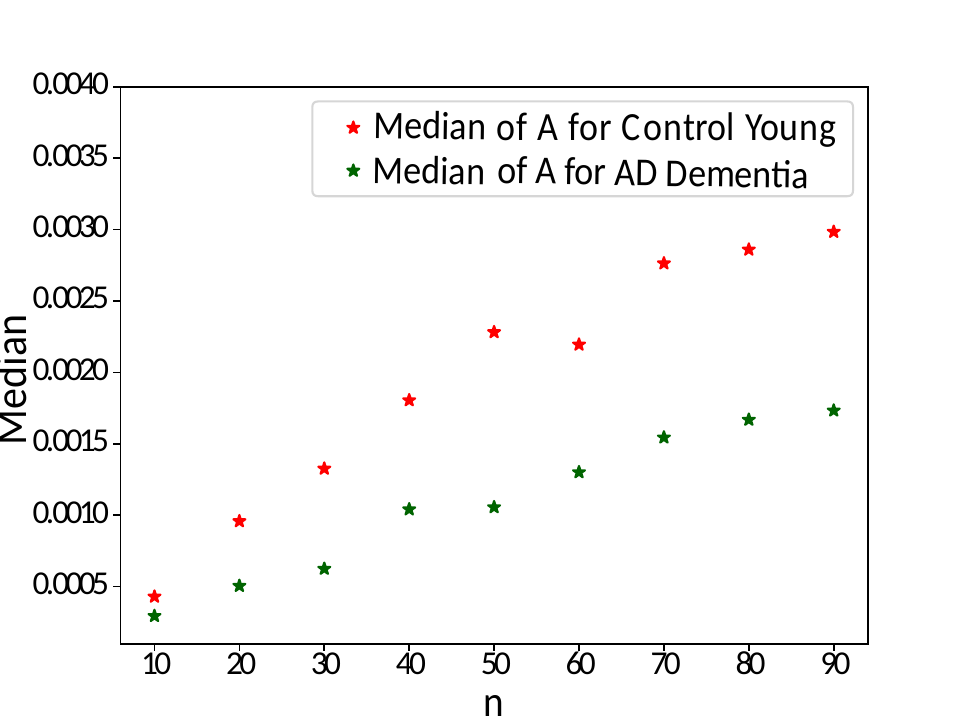}
    \includegraphics[width=0.32\linewidth]{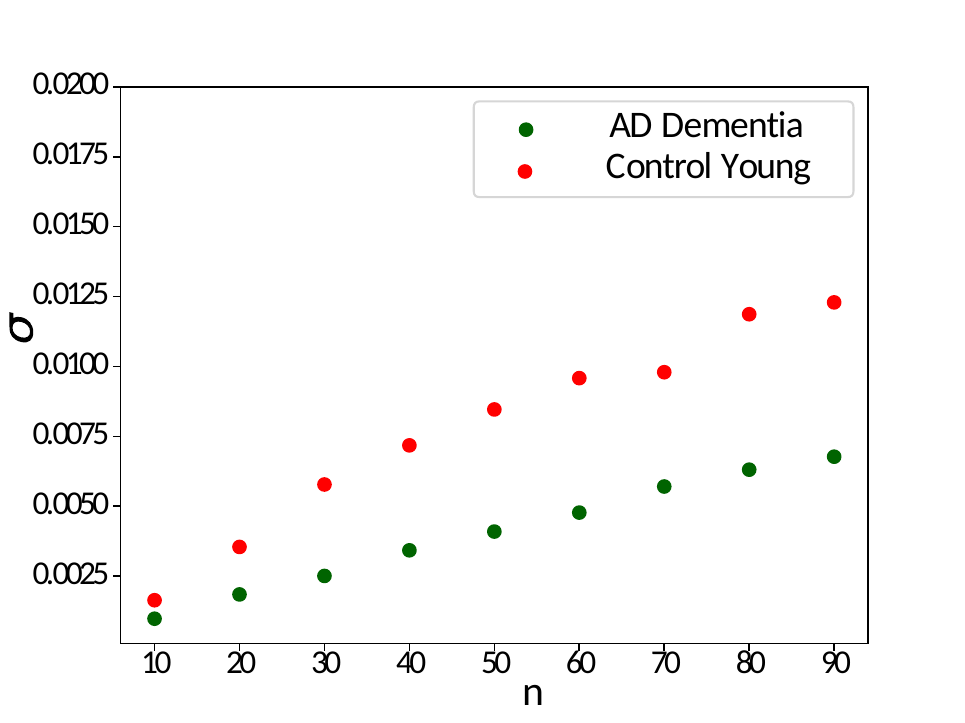}
    \caption{(Left) Frequency histogram of $\texttt{A}(n = 20)$, extracted from the network trajectory of a control young individual (red bars) and a patient diagnosed with AD Dementia (green bars).
    (Middle) Median of the population of $\texttt{A}$ as a function of the monitoring time $n$, comparing a control young individual (red points) to a patient diagnosed with AD Dementia (green points). The control individual systematically exhibits a larger median, suggesting that this metric may serve as a distinguishing parameter between cases.
    (Right) Same as the middle panel, but using the standard deviation $\sigma(\texttt{A}(n))$ of the population as the metric. }
    \label{fig:example_nonparametric}
\end{figure}

\medskip \noindent The algorithm we use to  measure $\texttt{A}(n)$ is as follows: first, we set the initial condition of interest, which is a point in the network trajectory (i.e. a network snapshot). Then, we inspect the network trajectory to find a nearby point (a recurrence in graph space), i.e., a sufficiently distant (in time) network snapshot which is sufficiently close (in graph space) to the initial condition. Specifically, we choose the closest in space to the initial condition, at a time distance of, at least, 250 time steps (corresponding to 500 milliseconds), to ensure that the time correlation between the two points has vanished. Note that this avoids choosing close-in-time snapshots as these are not recurrences.
We then track the distance between these two networks up to $n$ steps and compute $\texttt{A}(n)$ using Eq.~(\ref{eq:non_parametric}). Subsequently, we choose another initial network condition that is $n+1$ temporal steps away from the previous one, and repeat the procedure. Therefore, if $T$ is the total length of the trajectory, we will have a total population of $T/n$ estimates of $\texttt{A}(n)$. From this, a distribution of $\texttt{A}(n)$ can be obtained; examples for one young control participant and one AD-D patient are shown in the left panel of Fig.~\ref{fig:example_nonparametric}.

We then extract the median and the standard deviation from these histograms. A larger median of $\texttt{A}(n)$ indicates that  the subject displays faster expansion of the trajectories, a proxy for a less predictable brain dynamics. The standard deviation of the distribution of $\texttt{A}(n)$ for any one patient indicates the variability of predictability across time. The data in Fig.~\ref{fig:example_nonparametric} indicate that these summary statistics have the potential to discriminate between groups of participants. The right-hand panel shows the standard deviation $\sigma$ of $\texttt{A}(n)$ against the monitoring time $n$. As shown in the right panel, the standard deviation $\sigma(\texttt{A}(n))$ is systematically smaller for the control than for the patient, and such difference is consistent and is further amplified as the monitoring time $n$ increases. While $n$ could then be chosen to be arbitrarily large, a relatively small value of $n$ allows us to have more $\texttt{A}(n)$ points, resulting in better statistics. Consequently, in what follows, and unless otherwise specified, we set $n = 20$, equivalent to $40$ milliseconds. For the sake of clarity, we will simply write $\texttt{A}$ from now on.

\section{Results}

\begin{figure}[!tb]
\begin{center}
\includegraphics[width=0.99\linewidth]{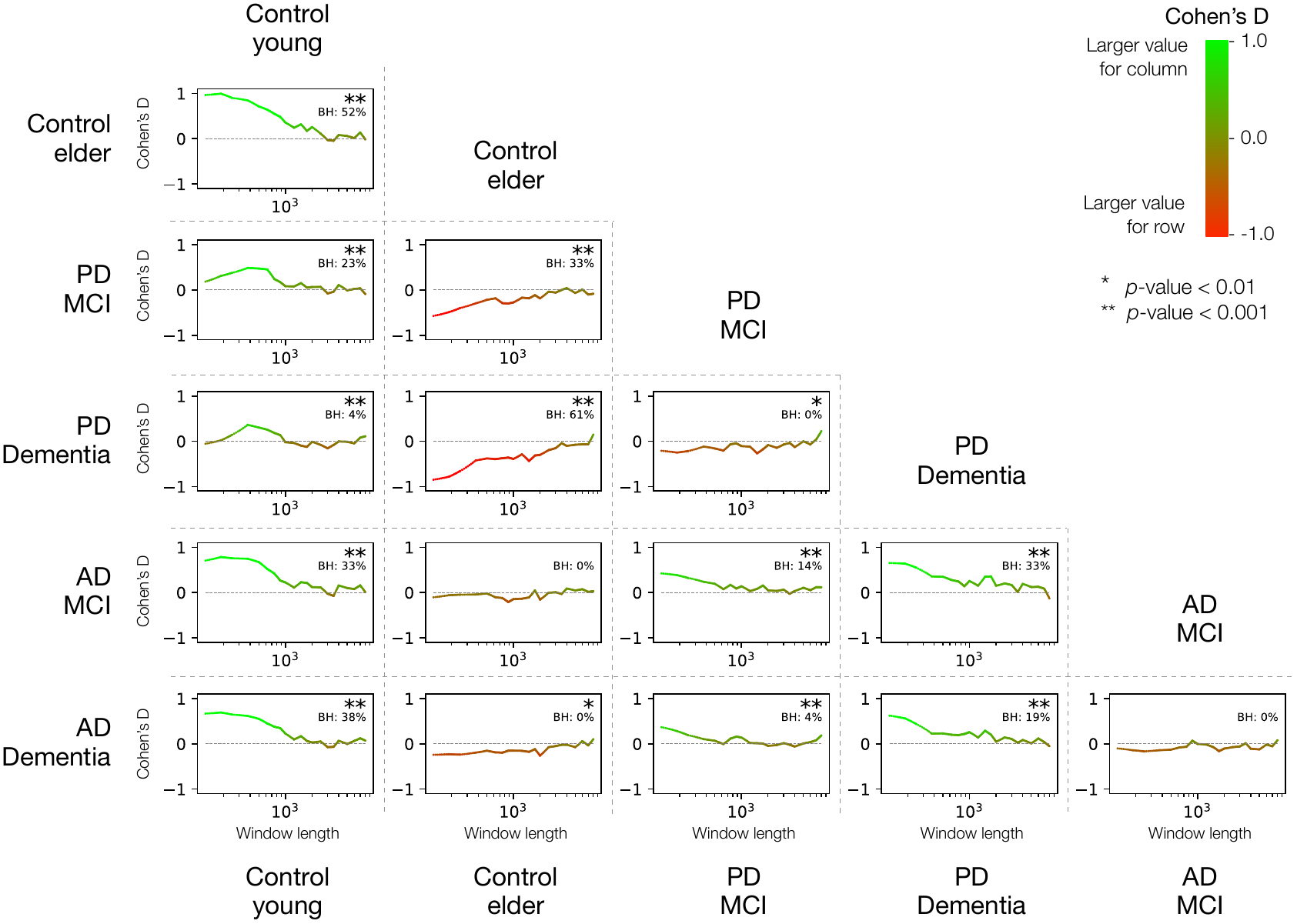}
\end{center}
\caption{Differences of the median of $\texttt{A}$ between groups. Each panel reports the evolution of the Cohen's $D$, calculated between the median of the distributions of $\texttt{A}$ of two groups, as a function of the window length.
Green points indicate positive $D$s, and hence that $\texttt{A}$ is higher for the group marked in the column; red points, negative $D$s. The $^{*}$ and $^{**}$ symbols indicate that the difference is statistically significant at levels $0.01$ or $0.001$ for at least one value of the window length. Each sub-plot further reports the percentage of window lengths for which a statistically significant difference is observed after a Benjamini-Hochberg (BH) correction. }\label{fig:MedianA}
\end{figure}

\subsection{Differences in the median of $A$ across groups}

We start the analysis by evaluating the differences in the median of $\texttt{A}$ across groups. Specifically, Fig. \ref{fig:MedianA} reports the difference between the median of $\texttt{A}$ as a function of the window length, i.e. the length of the time interval considered to compute the correlation between pairs of time series, and hence to reconstruct the functional networks. More specifically, for each trial, the distribution of $\texttt{A}$ is calculated (see left panel of Fig. \ref{fig:example_nonparametric}), and from it the median is extracted; next, given two groups of subjects, the difference between both sets of medians is encoded through the standardised effect size Cohen's $D$ metric \cite{cohen2013statistical}. 
In Fig. \ref{fig:MedianA}, the first group corresponds to the one in the column and the second to the one in the row. A positive Cohen's $D$ indicates that the column group has a higher median compared to the row group, while a negative value implies the opposite.
For a group to have a higher median than the other, it means that the former exhibits lower predictability.

It can be appreciated that, throughout most pairs, differences appear for short window sizes, usually below $10^3$ points (i.e. two seconds).
Some relationships are easy to describe; for instance, control young people always show a higher median than the other groups (positive Cohen's D metric corresponding to the control young column in Fig. \ref{fig:MedianA}) and exhibit the fastest expansions of the dynamics.
In other words, networks of control young participants evolve faster, and hence have the lowest predictability. Control elders present a more complicated scenario, in which the functional network dynamics expands slower than for PD MCI and PD Dementia (see the negative Cohen's D),
but is nevertheless qualitatively similar to AD. This, in turn, means that PD patients have a more unpredictable dynamics than matched healthy subjects; while AD patients have a similar one. 

Note that, in most pairs of subject groups, it is possible to find a statistically significant (with significance level $\alpha < 0.001$) difference between them for at least one value of the window length - as indicated by the $^{**}$ symbol.
This may nevertheless be caused by the large number of window lengths tested in each graph. To complement this information, each sub-plot of Fig. \ref{fig:MedianA} reports the percentage of window lengths for which a statistically significant difference is observed, after a Benjamini-Hochberg (BH) procedure for controlling for false discovery rate \cite{wilcox2011introduction}. Note that values higher than $30\%$ are achieved for many pairs, supporting the statistical significance of the differences here observed.

\begin{figure}[!tb]
\begin{center}
\includegraphics[width=0.99\linewidth]{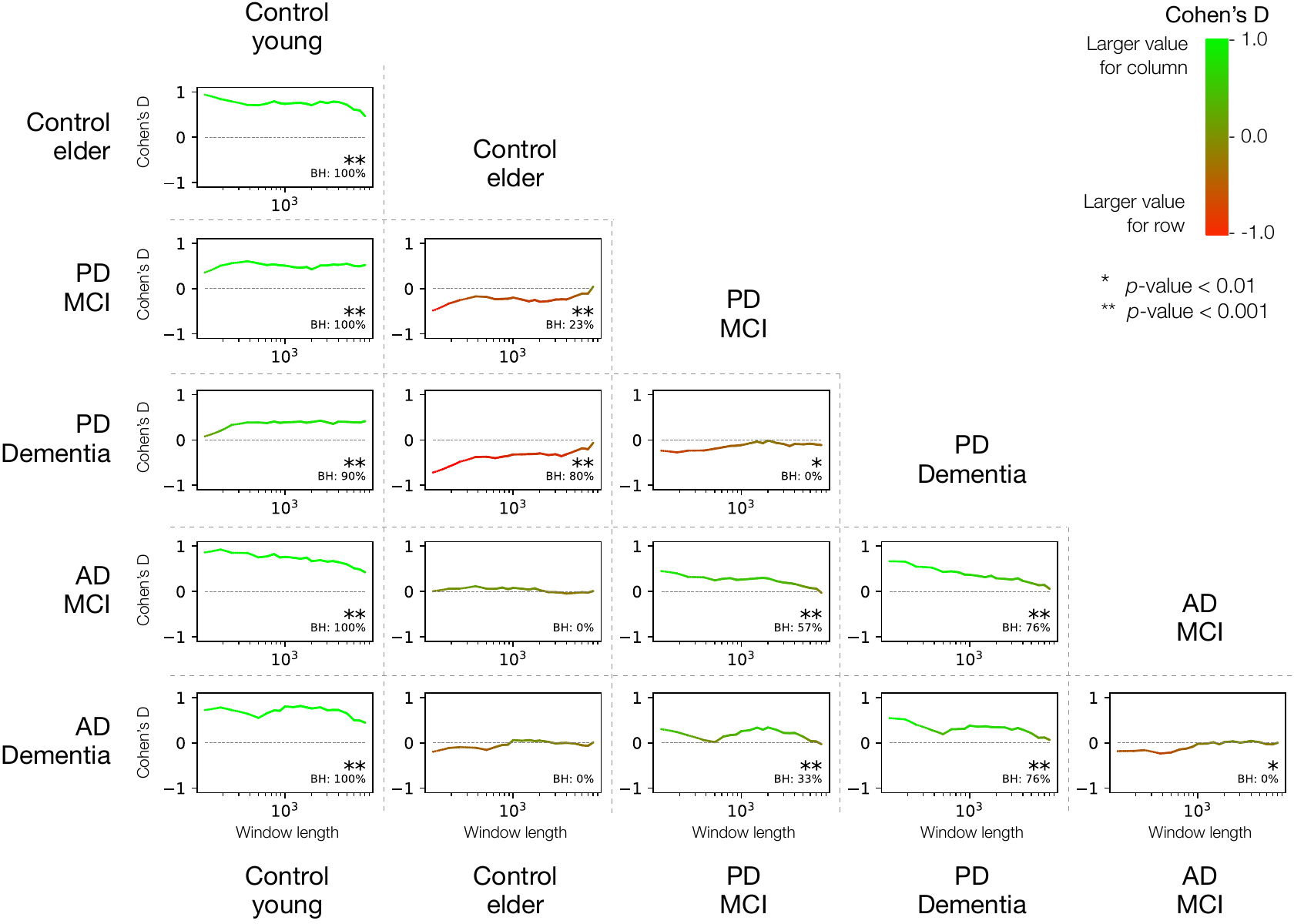}
\end{center}
\caption{Differences of the $\sigma(\texttt{A})$ between groups. Each panel reports the Cohen's $D$ (which evaluates here the difference between pairs of groups based on the metric $\sigma(\texttt{A})$), as a function of the window length over which each functional network snapshot is calculated. Green points ($D>0$) indicate that the group marked in the column has larger $\sigma(\texttt{A})$; the opposite ($D<0$) is true for red points. The $^{*}$ and $^{**}$ symbols indicate that such difference is statistically significant at levels $0.01$ or $0.001$ for at least one value of the window length, respectively. Each sub-plot further reports the percentage of window lengths for which a statistically significant difference is observed after a Benjamini-Hochberg (BH) correction. }\label{fig:SigmaA}
\end{figure}

\subsection{Differences in the standard deviation of $A$ across groups}

We then move to the analysis based on the second pointwise estimator, $\sigma(\texttt{A})$, which represents the spread of the distribution. To assess differences in this spread across groups, we apply Cohen's D directly to the standard deviations, treating them as the quantities of interest. These differences are analysed as a function of the window length — see Fig. \ref{fig:SigmaA}. In this section, a positive Cohen's D indicates that the group in the column has a higher average standard deviation, implying that \texttt{A}s are more spread out, and hence higher fluctuations in the predictability for this group.

It can be appreciated that results are very similar to what previously observed in Fig. \ref{fig:MedianA}, with young control subjects having the least consistent predictability (i.e. positive D in the first column); PD patients being less consistent than matched controls (negative D in second column), but yet, slower than young ones; and with AD patients being qualitatively similar to matched controls. The main notable point with respect to Fig. \ref{fig:MedianA} resides in the fact that differences appear at all window lengths, and not only at short ones - see the generally high values of BH significance percentages.

\begin{figure}[!tb]
\begin{center}
\includegraphics[width=0.99\linewidth]{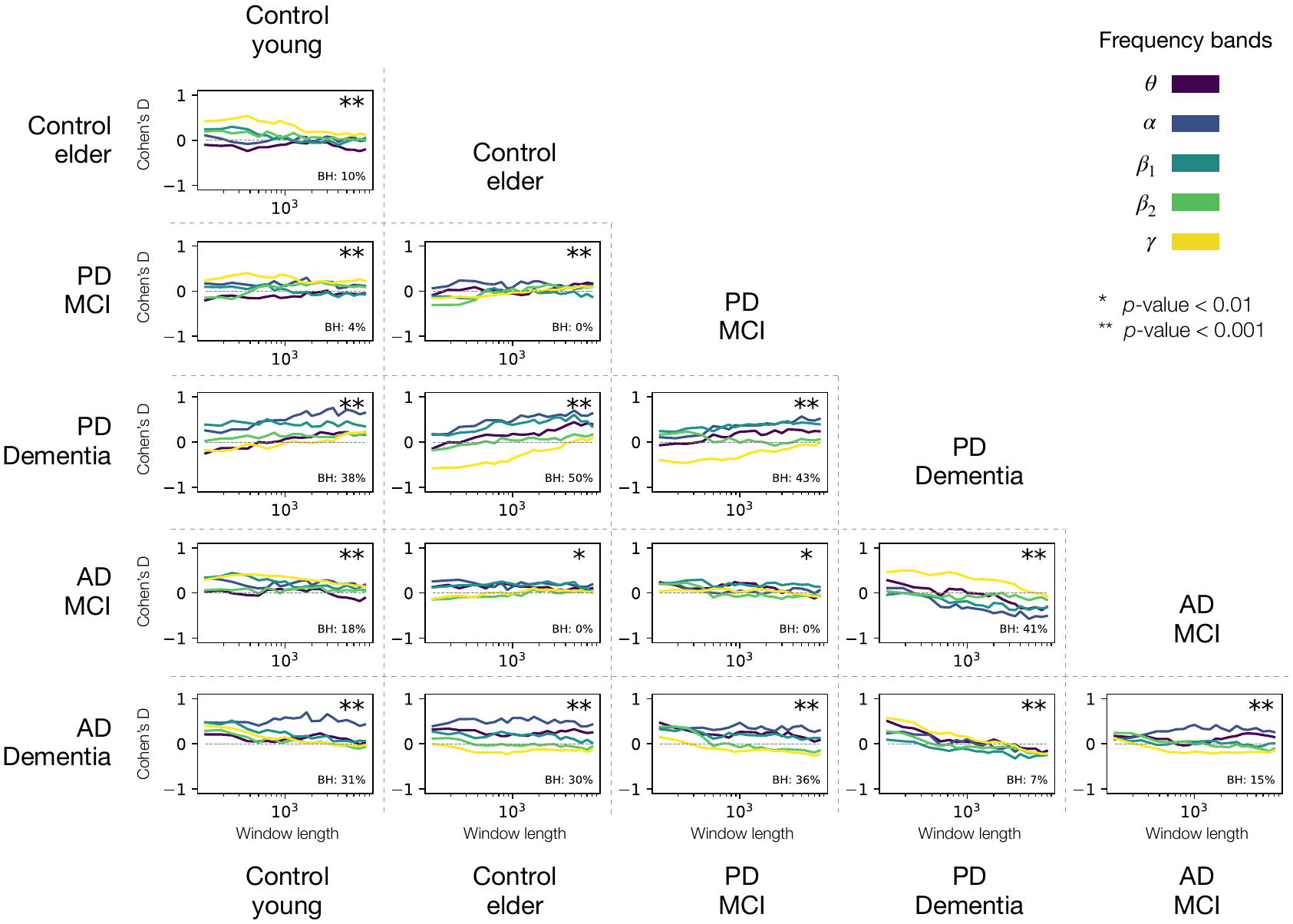}
\end{center}
\caption{Differences in $\sigma(\texttt{A})$ by frequency bands. Each panel reports the evolution of the Cohen's $D$, calculated between the distributions of the $\sigma(\texttt{A})$ in two groups, as a function of the window length and the frequency band (see right panel). Positive and negative values of $D$ have the same meaning as in Fig. \ref{fig:SigmaA}. The $^{*}$ and $^{**}$ symbols indicate that such difference is statistically significant at levels $0.01$ or $0.001$ for at least one value of the window length and one frequency band, respectively. Each sub-plot further reports the percentage of tests for which a statistically significant difference is observed after a Benjamini-Hochberg (BH) correction. }\label{fig:Freq}
\end{figure}

\subsection{Breakdown into frequency bands}

In order to assess whether these patterns hold at different frequencies, we have performed the same computation of the evolution of $\sigma(\texttt{A})$ by filtering the EEG time series according to five standard frequency bands: $\theta$ ($4-8$ Hz), $\alpha$ ($8-13$ Hz), $\beta_1$ ($13-20$ Hz), $\beta_2$ ($20-30$ Hz), and $\gamma$ ($30-50$ Hz). Fig. \ref{fig:Freq} then reports the same results as in Fig. \ref{fig:SigmaA}, but with one line for each frequency band (see right legend).
Differences between groups in each frequency band are mostly constant across window lengths; or, in other words, the internal ranking of the bands does not substantially change when changing the window length.
$\gamma$ and $\alpha$ bands (respectively yellow and dark blue lines) seem to mostly be at the extreme, i.e., they define the largest and lowest values of $D$; in many cases, a positive value of $D$ for one of them implies a negative D for the other.
Other frequency bands, and most notably $\theta$, have a more intermediate role, with $D \approx 0$ and hence minimal differences across groups.

\begin{figure}[!tb]
\begin{center}
\includegraphics[width=0.99\linewidth]{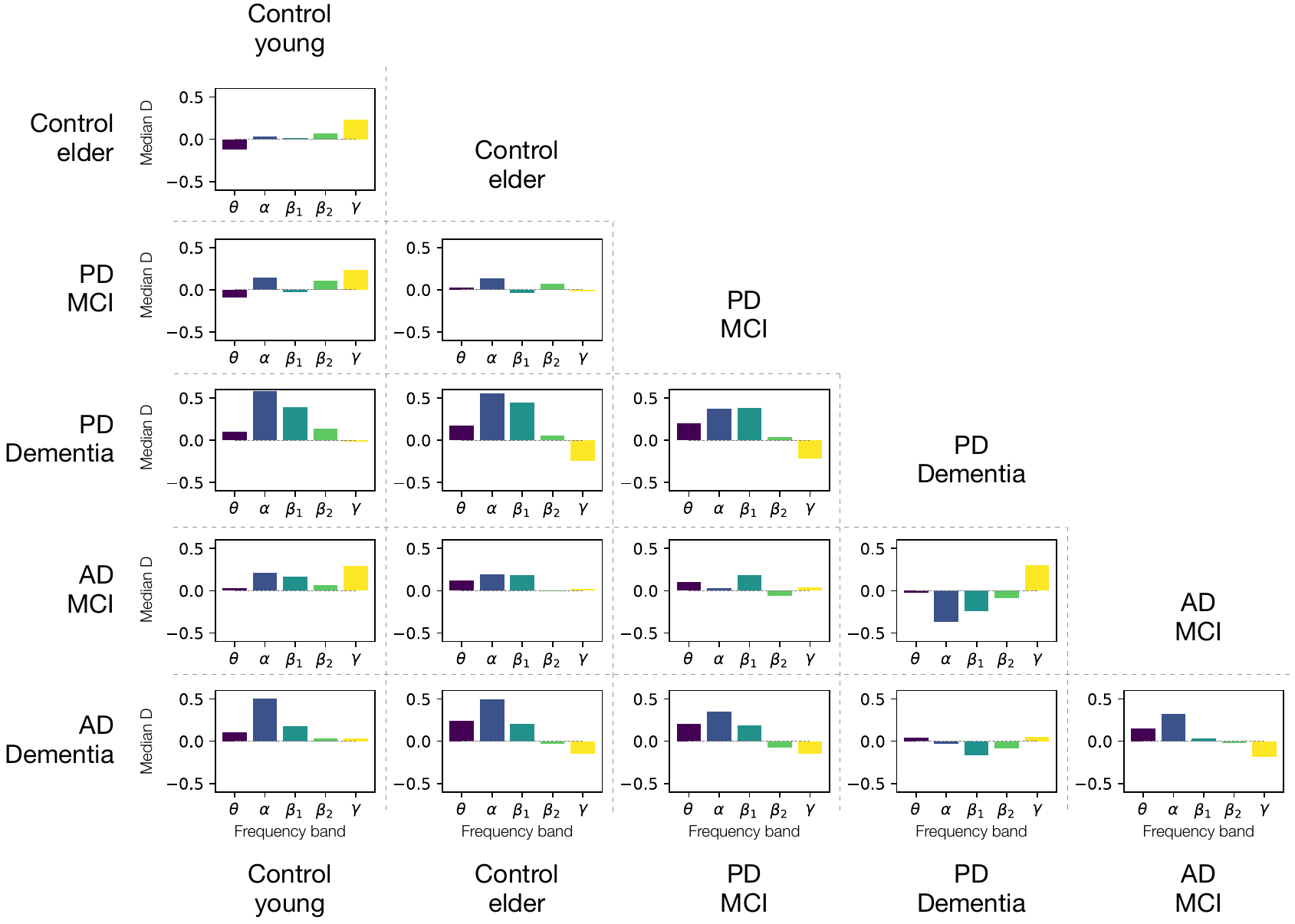}
\end{center}
\caption{Median differences in $\sigma(\texttt{A})$ by frequency bands. Each panel reports the median of Cohen's $D$, which is calculated for $\sigma(\texttt{A})$, across all values of the window length and for different frequency bands, i.e., the median of each plot in Fig. \ref{fig:Freq}.
Positive and negative values of $D$ have the same meaning as in Fig. \ref{fig:SigmaA}.}\label{fig:FreqMedian}
\end{figure}

In order to simplify the interpretation of the previous figure, Fig. \ref{fig:FreqMedian} shows a synthesis of the same results. Specifically, for each pair of groups and frequency band, it shows the median value of the Cohen's $D$ across all considered window lengths.

\section{Discussion and conclusions}

Brain activity presents non-random spatiotemporal structure and is naturally modelled as a high-dimensional networked dynamical system operating at multiple scales. Simultaneously accounting for the spatial and temporal aspects of such a structure is therefore non-trivial, and characterisations of brain activity often concentrate on one aspect, approximating the other. In this study, we reconstruct network trajectories \cite{lacasa2022correlations} of electrical brain activity, we introduce a non-parametric method to quantify the predictability of network trajectories --which generalises network extensions of Lyapunov exponents \cite{caligiuri2023lyapunov}-- and apply it to characterise spontaneous electrical brain activity in various patient populations as well as in young and elderly healthy control subjects. Our results generally show that this predictability measure significantly differs across these populations and across timescales in a stylised manner. Let us now discuss in some detail these results and their implications, starting from the differences across populations, for then moving to frequencies and time scales.

\medskip
Overall, fully-fledged dementia appears to be associated with increased median predictability (as compared  to young healthy subjects) at short time scales, and with decreased predictability fluctuations at all scales (Figs. \ref{fig:MedianA} and \ref{fig:SigmaA}). This interpretation is consistent with the one found in a previous study that followed a different methodology \cite{zanin2022telling}, where pathologies were found to be associated with lower stability of links, but with higher persistence of larger-scale features (e.g. node centrality and clustering) and, more generally, with a picture wherein essentially normal function was maintained by compensating the slight decrease in connectivity and increased hub persistence via an overall slower dynamics. These works reflect on the idea that intrinsic variability is a sign of the adaptability of the underlying control networks and therefore of a healthy system \cite{west1987physiology, goldberger2002fractal}.

\medskip \noindent 
Insofar as our proposed measure quantifies the predictability of network dynamics --and therefore speaks of the connectivity induced by brain activity--, it is interesting to compare these results to previous studies on synchronisation in AD and in its harbinger MCI.  Hypersynchrony has been suggested to be associated with MCI \cite{stam2003eeg, fide2022hyperconnectivity}, particularly for subjects eventually transitioning to dementia \cite{pusil2019hypersynchronization}. Such an activity would represent a dynamic compensation of the structural damage, preceding the network breakdown characterising fully-fledged AD dementia \cite{stam2003eeg}.

\medskip \noindent 
However, there are important differences in short term predictability between the paths to AD and PD (Figs. \ref{fig:MedianA} and \ref{fig:SigmaA}). In the path to AD dementia, predictability is found to increase progressively (i.e. uncertainty decreases) from young healthy subjects to fully-fledged dementia. Various aspects are worth highlighting. First, while the median predictability tends to differ between young healthy controls and all other conditions from healthy ageing to fully-fledged AD at short but not at long timescales, the standard deviation of such measure differs essentially at all scales. Second, healthy ageing does not appear to be distinguishable from MCI in terms of short-term predictability. However, healthy ageing significantly differs in median predictability from fully-fledged AD, though differences are much less pronounced than those with young healthy subjects, but not in predictability's standard deviation, whereas the opposite pattern characterises differences between MCI and fully-fledged AD, as conditions only differ in terms of predictability's standard deviation.  

\medskip \noindent 
Predictability also changes in a progressive fashion in the path to PD dementia. Predictability is always lower for the young control group, i.e. dynamical network structure evolves faster than in any other group, while fully-fledged PD dementia appears to be the end point of a continuous increase of predictability from healthy ageing to dementia (see Figs. \ref{fig:MedianA} and \ref{fig:SigmaA}). However, there is an important difference with respect to predictability in AD: spatiotemporal brain activity in healthy ageing appears to be significantly more predictable than in both MCI and in fully-fledged PD dementia, particularly for short time-windows, PD patients with dementia turning out to be the least predictable.

\medskip \noindent 
Overall, our results suggest that, in addition to what is known to be a progressive tendency to a random but topologically homogeneous spatial structure \cite{buldu2011reorganization}, the dynamical path to dementias is associated with a progressively more regular --i.e., more predictable-- temporal structure.

\medskip
The role of timescales can be appreciated by looking at how our metrics varies (i) as a function of the window length within which networks were reconstructed, and (ii) for different frequency bands (see Fig. \ref{fig:Freq}). Both in AD and PD dementia, predictability varies with the size of the time-window used to reconstruct brain networks. On the other hand, frequency-specific analyses (see Figs. \ref{fig:Freq} and \ref{fig:FreqMedian}), show that AD Dementia is characterised by higher predictability in the $\alpha$ and $\beta_1$ bands, compared to both healthy control groups and AD MCI patients.
Interestingly, the predictability of $\gamma$ band activity increases from young control subjects to elderly control, PD-MCI, and AD-MCI subjects, but does not in PD and AD dementia (see yellow bars in the first column of Fig. \ref{fig:FreqMedian}). 
PD dementia is characterised by similar though more pronounced increases in $\alpha$ and $\beta_1$ band predictability and by decreases in $\gamma$ band predictability.

These results are somehow coherent and complementary with previous results concerning the frequency composition of brain activity in AD and PD. EEG activity in AD is typically characterised by a progressive slowing of brain oscillatory activity \cite{jeong2004eeg, dauwels2011slowing, hsiao2013altered, ishii2018healthy}, with increased power in the $\delta$ and $\theta$ frequency bands, mirrored by decreased power in the $\alpha$ and $\beta$ (12–30 Hz) bands, as well as complex changes in the $\gamma$ band \cite{roh2011region, jafari2020neural, guntekin2022there, guntekin2023event}, a pattern already visible prior to fully-fledged AD \cite{babiloni2004mapping, babiloni2009hippocampal, babiloni2010cortical, dauwels2011slowing, jelic2000quantitative, lopez2016alpha, bruna2023meg}. On the other hand, PD patients' EEG activity generally presents higher $\theta$ and $\delta$ band power \cite{han2013investigation}, as well as lower power and decreased dynamical connectivity in the $\beta_2$ band compared to cognitively unimpaired PD patients, and reduced power in the $\gamma$ band in central, parietal, and temporal regions. Compared with early-stage PD patients, late-stage PD patients have been associated with reduced power in the $\beta_1$ range in the posterior central region, and increased $\theta$ and $\delta$ power in the left parietal region \cite{chang2022evaluating}.
Moreover, significant differences in EEG activity was found between patients with PD and AD patients, with more pronounced slowing in PD patients' activity compared to AD patients \cite{benz2014slowing}.

Notably, somehow counterintuitively, window length and frequency bands are not simply related to each other, and the Cohen's D metric for $\sigma(\texttt{A})$ across frequency bands remains constant as the window length increases (see Fig. \ref{fig:Freq}). Window lengths and frequency bands are therefore capturing two different aspects, though why this is the case and what exactly their respective role is not entirely clear.

\begin{figure}[!tb]
\begin{center}
\includegraphics[width=0.99\linewidth]{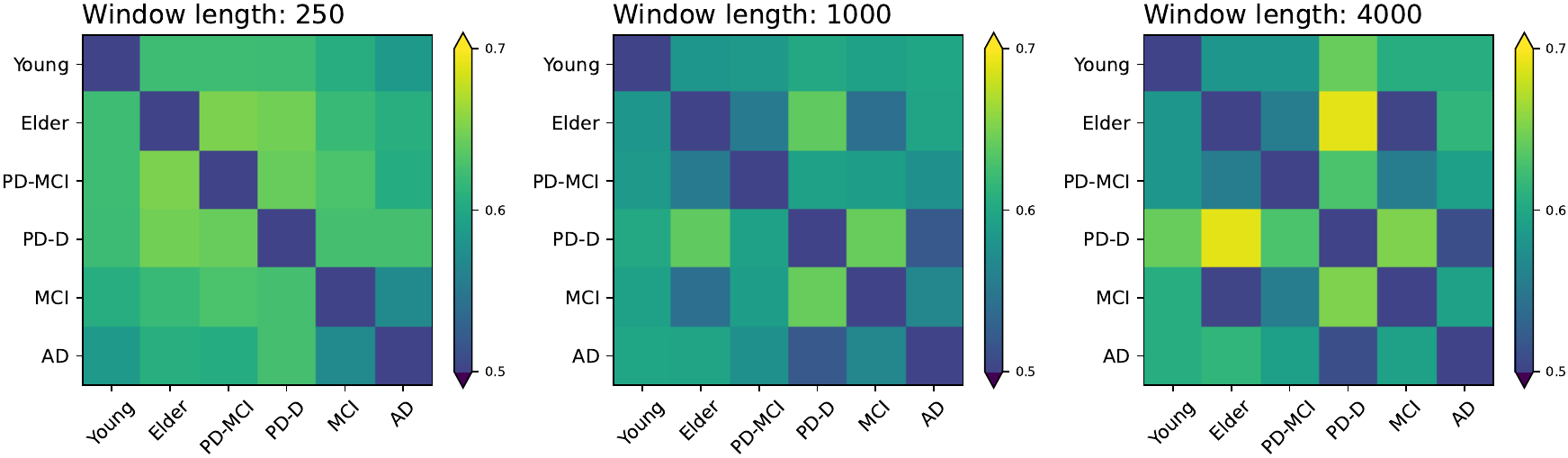}
\end{center}
\caption{ Classification score obtained by Random Forest machine learning models while pairwise discriminating the participants' groups. Each model has been trained with the $\sigma(\texttt{A})$ obtained from each trial for the five frequency bands considered in this study (see Fig. \ref{fig:Freq}); the three panels report the results for different window lengths (see top title). All results correspond to the average accuracy obtained over $100$ independent realisations. }\label{fig:ClassScore}
\end{figure}

\medskip
To round off the analysis, and since most of the differences between groups are highly statistically significant, it may be tempting to expect that such differences can completely {\it define} the groups, i.e. that a model on them based could be used to classify or diagnose subjects. We tested this idea by resorting to a random forest machine learning model \cite{breiman2001random}, trained on the $\sigma(\texttt{A})$ values for the five frequency bands (see Fig. \ref{fig:FreqMedian}), and aimed at pairwise discriminating the participants' groups. Results are measured through the accuracy obtained in each classification task, i.e. the fraction of participants whose group has correctly been classified; and their generalisability is further evaluated using a ten-fold cross-validation procedure \cite{browne2000cross}. In all tasks, a same number of randomly-selected trials is considered for both groups, to ensure balanced data. Results are reported in Fig. \ref{fig:ClassScore}, for three window lengths and for each pair of groups. The obtained classification scores are generally not very high, with a maximum of $0.69$
for the task control elder vs. PD Dementia (to be compared to $0.5$ that would be expected in a random classification). In other words, in most cases the classification is only moderalety better than what would be obtained by chance. The differences here presented thus do not provide a mean to exactly {\it separate} groups, and have a more descriptive value, highlighting alterations in brain's dynamics which cannot buttress predictive tasks by themselves.

\medskip 
As a final note, it is worth highlighting that the analyses here presented are based on two main methodological choices: (i) linear correlations have been used to reconstruct functional networks, among the many alternatives available in the literature \cite{jalili2016functional}, and (ii) a concrete network distance function (Eq.~\ref{eq:distance}), among other choices \cite{wills2020metrics}. While these choices were guided by parsimony, we cannot rule out that the use of alternative metrics may elucidate further aspects of the predictability of brain dynamics, as e.g. their connection with the non-linear coupling between brain regions.

\backmatter

\section*{Declarations}

\subsection*{Funding}
This project has received funding from the European Research Council (ERC) under the European Union's Horizon 2020 research and innovation programme (grant agreement No 851255). 
The authors further acknowledge partial support from projects MISLAND (PID2020-114324GB-C22) and the Mar\'ia de Maeztu project CEX2021-001164-M, funded by MICIU/AEI/10.13039/501100011033. 
AC acknowledges funding by the Maria de Maeztu Programme (MDM-2017-0711) and the AEI (MICIU/AEI/10.13039/501100011033) under the FPI programme. 

\subsection*{Competing interests}
The authors declare that the research was conducted in the absence of any commercial or financial relationships that could be construed as a potential conflict of interest.

\bibliography{BrainNetLE}

\end{document}